\documentstyle[12pt,psfig,aaspp]{article}
\def\beq{\begin{equation}}
\def\eeq{\end{equation}}

\def\mol{$M/L~$}
\def\av#1{\langle #1\rangle}
\def\cmss{cm s$^{-2}$}
\def\tot{{2\over 3}}
\def\slos{\sigma_{los}}
\def\a0{a_0}

\def\vv{{\bf v}}
\begin{document}

\title{MOND mass-to-light ratios for galaxy groups}
\author{Mordehai Milgrom }
\affil{ Department of Condensed Matter Physics Weizmann Institute}

\begin{abstract}

I estimate MOND \mol values for nine galaxy groups that were
recently studied by Tully et al.. Instead of the large \mol values
that they find with Newtonian dynamics (up to 1200 solar units)
the MOND estimates fall around 1 solar unit. Tully et al. find a
systematic and significant difference between the \mol values of
groups that do not contain luminous galaxies and those that do:
Dwarfs-only groups have larger \mol values (by a factor of $\sim
5$). The MOND \mol values do not show this trend; the Newtonian
disparity is traced back to the dwarfs-only groups having
systematically smaller intrinsic accelerations (similar sizes, but
rather smaller velocity dispersions).
\end{abstract}
\keywords{Galaxies: kinematics and dynamics; Cosmology: dark
matter}

\section{introduction}
\par
It is important to test MOND (modified Newtonian dynamics) on
systems other then disc galaxies, for which it arguably performs
well (see, e.g., \cite{sanrcs,dbm,sv}, and, for a recent review on
MOND, \cite{araa}). And, it has been pointed out that MOND does
not fully explain away the mass discrepancy in the inner parts of
x-ray galaxy clusters (\cite{the,gerbal,aguirre}). Loose galaxy
groups have masses similar to those of galaxies (or somewhat
larger) and sizes comparable with those of the inner regions of
galaxy clusters (mean projected radii of several hundred kpc); we
thus note, in the context of MOND, that galaxy groups probe rather
smaller accelerations than either galaxies or clusters: The
typical accelerations in such groups are an order of magnitude
smaller than those reached in LSB galaxies, or in the outskirts of
HSB galaxies.
\par
MOND analysis of groups has so far been applied only to mean
properties of whole group catalogues. Based on published mean
values of luminosities and velocity dispersions, \cite{groups}
estimated MOND mean-mass-to-mean-luminosity values for four group
catalogues. Values of a few solar units were found, instead of
100-200 solar units found in a Newtonian analysis.
\par
 \cite{tully} have recently published data and analysis
 for nine individual, nearby groups. These, with the group
 parameters listed by them, lend themselves to MOND analysis.

 This small new sample is particularly interesting in the present context
 because it lists separately groups that contain luminous galaxies and those
 comprising only dwarf galaxies, and because \cite{tully} find that the
  latter have
   \mol values of $\sim300-1200$ solar units, significantly larger
 than those of the former, with $\sim10-150$ solar units.
In MOND, large mass
 discrepancies are supposed to bespeak low accelerations; so,
 this dichotomy should follow from a disparity in the
 characteristic accelerations in these two types of groups.
 Inasmuch as the \mol values for individual groups are still rather
  uncertain, this affords an interesting
 statistical test,
 intermediate between
 testing individual groups and testing mean values for the
 whole sample.

\section{METHOD}
\par
I use two MOND mass estimators for the groups

\beq M\approx {81\over 4}\slos^4(G\a0)^{-1}, \label{nusa} \eeq and
\beq M\approx {81\over 4}\slos^4(G\a0)^{-1}(1-N^{-1/2})^{-2},
\label{nusa1} \eeq
 where $\slos$ is the line-of-sight velocity
dispersion, $N$ is the number of galaxies listed for the group,
and $\a0$ is the MOND acceleration constant taken to be
$\a0=1.2\times 10^{-8}$\cmss, as deduced from the rotation-curve
analysis of \cite{bbs}. These estimators can be derived as
approximations to the relation
\beq\av{\av{(\vv-\vv_{com})^2}}_t=\tot(MG\a0)^{1/2}[1-\sum_i
(m_i/M)^{3/2}], \label{nus} \eeq
 where $\vv$ is the 3-D velocity,
$\vv_{com}$ is the center-of-mass velocity,
 $\av{}$  is the mass-weighted average over the constituents, whose
masses are $m_i$, $\av{}_t$ is the long-time average, and $M$ is
the total mass.
\par
  Relation(\ref{nus}) (\cite{lsy,conf}) is exact for a
  bound system of point masses
  in the formulation of MOND as modified gravity (\cite{bm})
in the deep-MOND limit: accelerations much smaller than $\a0$. It
is also assumed that the system is isolated in the MOND sense,
i.e., is not subject to an external field. (Interestingly, the
fact that the time-average rms velocity dispersion depends solely
on the constituent masses--and not, e.g., on system size--follows
from the conformal invariance of this limit of the theory, as
shown by \cite{conf}.) All groups in the \cite{tully} sample are,
indeed, deep in the MOND regime.
\par
The assumptions and approximations leading from relation
(\ref{nus}) to the simplified relations (\ref{nusa}) and
(\ref{nusa1}) are discussed in more detail in \cite{groups}.
Briefly, (i) I drop the long-time average, and (ii) replace the
3-D quantity $\av{(\vv-\vv_{com})^2}$
 by the line-of-sight, statistical substitute $3\av{(v-v_{com})^2_{los}}$.
 These assumptions are also, effectively, made
in the Newtonian analysis. Also, (iii) \cite{tully} give not the
mass-weighted velocity dispersion as needed in eq.(\ref{nus}),
 but the unweighted line-of-sight dispersion
 $\sigma_v$, which I use instead. And,
(iv) in eq.(\ref{nusa}) I approximate the right-hand side of
eq.(\ref{nus}) by $\tot(MG\a0)^{1/2}$, which is valid
 in the limit of large number of constituents, $N$, each having a mass
$\sim M/N\ll M$. In eq.(\ref{nusa1}) which gives a higher
estimate, I don't assume that $N$ is very large. I use
eq.(\ref{nusa}) as a useful value beside estimator (\ref{nusa1}),
because the latter
 is too large when the masses are not
equal, and because $N$ given in \cite{tully} is not the full
number of galaxies in the group, which should be used in the
estimator. For example, in the case of one very dominant galaxy,
with all the rest being of equal and negligible masses, even the
smaller estimate (\ref{nusa}) is too large and has to be reduced
by a factor $4/9$.
\par
 As explained in \cite{groups}, the possible breakdown of these
assumptions introduces, typically, `factor-of-a-few' errors.
Furthermore, it is hardly ever certain, for an individual group
candidate, that the assumptions underlying eq.(\ref{nus}) itself
hold. The questions of contamination by interlopers, boundedness
of the group, and virial equilibrium always loom, and can
introduce large errors. In MOND, there is an additional worry
having to do with the external-field effect (EFE): If the group is
falling in the field of an external structure with an
acceleration, $a_{ex}$, larger than its internal accelerations,
$a_{in}$, then eqs.(\ref{nusa})-(\ref{nus}) do not apply. These
expressions then underestimate the mass, and \mol value, of the
group by roughly $a_{in}/a_{ex}<1$.
\par
The issue of virialization is particularly worrisome in light of
the large dynamical times for some of the groups in the sample.
However, our estimators are useful even if the dynamical time is
comparable with the Hubble time, provided it is also comparable
with the lifetime of the group. This is because, by MOND, the
typical acceleration with which the system is collapsing is
$g\sim(MGa_0/R^2)^{1/2}$ ($R$ a characteristic radius of the
group). And, if the collapse time can be approximated by $\tau\sim
R/v$, with $v$ the representative three-D velocity, then $v\sim
gt\sim (MGa_0)^{1/2}/v$, from which eq.(\ref{nusa})(\ref{nusa1})
follow as order-of-magnitude approximations.

\section{Results and discussion}

\par
The MOND \mol estimates for the nine groups are presented in the
last two columns of Table 1 together with the Newtonian values
(column 7) and other pertinent group parameters from \cite{tully}:
group designation (c. 1), number of galaxies included (c. 2), mean
projected radius $R_p$ (c. 3), velocity dispersion $\sigma_v$(c.
4), $\tau=R_p/\sigma_v$ as some measure of the dynamical time (c.
5), and total luminosity (c. 6). The last four lines are for
dwarfs-only groups.
\par
We see that the MOND \mol estimates fall around a few solar units.
The group 14+13 is the only one with an unacceptably small MOND
\mol value (see below).
\par
 We also see no
systematic difference between the dwarf-only groups and those
containing luminous galaxies. The large disparity in their
Newtonian \mol values is traced back to the significantly smaller
acceleration in the dwarf-only groups. These have similar radii to
those of the luminous groups but rather smaller velocity
dispersions.
\par
Some of the groups have $\tau$ values of order, and even
exceeding, the Hubble time. The group 14+13 has a particularly
long dynamical time, which perhaps explains the too low value of
\mol in MOND. Its listed velocity dispersion (12 km s$^{-1}$) is
exceptionally small compared with the other non-dwarf groups
(between 50 and 100 km s$^{-1}$) and may be far below the virial
velocity, not yet achieved. (To get a MOND \mol value of 1 solar
unit with estimator (\ref{nusa1}) the virial velocity dispersion
has to be 35 km s$^{-1}$.) Alternatively, we may be seeing a
virialized, quasi-planar, low-inclination system; in which case,
again, the observed line-of-sight dispersion is much below the
value that should go into the mass estimator. A possible
involvement of the EFE may also have to be reckoned with (see
below).
\par
It is difficult to estimate the external field in which individual
groups in the study are falling--due, say, to large-scale
structure--and so to assess the importance of the EFE. But, values
of order $0.01\a0$ are not unreasonable. For example, this is,
roughly, the MOND acceleration 130 Mpc away from a galaxy cluster
with an asymptotic, isotropic, line-of-sight velocity dispersion
of 1000 km s$^{-1}$ (30 Mpc for 500 km s$^{-1}$). Since the groups
under study are within 5 Mpc of us, even the Coma and Virgo
clusters, were they each the only attractor present, could
contribute accelerations of this magnitude. (Remember that when
there is more then one attractor the MOND acceleration is not the
sum of contributions.) The physics of the EFE implies that for an
external acceleration of $\eta\a0$ the MOND-to-Newtonian mass
ratio
 cannot be smaller than roughly $\eta$.
 The MOND-to-Newtonian mass ratios that I find for some of
 the groups [especially with estimator (\ref{nusa})] are comparable to,
  or smaller than $1/100$. This means that in some of the groups,
  the MOND mass estimates in Table 1 may be too
   small because they ignore the EFE.
\par
Note also that the \mol values in the table are not the stellar
 values, but the total (baryonic) ones. Because in some of these groups
the gas fraction is considerable, the \mol values have to be
corrected down to yield the stellar ones. For the 14+13 group,
 the MOND mass estimate using eq.(\ref{nusa1}), of $\sim
 10^8~M_\odot$, is
totally unacceptable, because it is smaller than even the gas mass
alone, deduced to be about $5\times 10^9~M_\odot$.

Clearly then, the individual MOND \mol values for the groups are
highly uncertain. All we can say is that MOND does correct the
huge Newtonian \mol values down to proportions compatible with
baryonic mass alone, and with no systematic differences apparent
between the two classes of groups.
\par
I thank the referee for very helpful suggestions.

\begin{table}
\caption{System parameters and $M/L$ values: Newtonian ($N$) and
two MOND estimates ($M$) for the groups. The last four are groups
comprising only dwarfs.}

\begin{tabular}{ccccccccc} \hline\hline (1) & (2) & (3) & (4) &
(5) & (6) & (7) & (8) & (9)\\ Group & No & $R_p$ &
 $\sigma_v$ & $\tau(R_p/\sigma_v)$ & $L$ & $(M/L)_N$ &
$(M/L)_M[eq.(\ref{nusa})]$ & $(M/L)_M[eq.(\ref{nusa1})]$\\
&& kpc & km~s$^{-1}$ & $10^{10}$y & $10^8 L_\odot$ &
$M_\odot/L_\odot$ & $M_\odot/L_\odot$ &
$M_\odot/L_\odot$ \\
\hline\hline
14-7  & 22 & 538 & 53 & 1.  & 264  &   72 & 0.38 & 0.6\\
14-10 & 12 & 322 & 107& 0.3 & 304  &  127 & 5.5  & 10.9\\
14-13 &  7 & 495 & 69 & 0.7 & 231  &   90 & 1.2  & 3.1\\
14+13 &  4 & 394 & 12 & 3.  & 72   &   13 &0.0036 & 0.015\\
14-12 & 16 & 178 & 77 & 0.2 & 409  &   50 & 1.1  & 2.0\\
\hline
14+12 &  6 & 569 & 22 & 2.  & 5.4  & 1220 & 0.55 & 1.6\\
14+8  &  3 & 180 & 16 & 1.  & 3.1  &  250 & 0.27 & 1.5\\
14+19 &  4 & 356 & 28 & 1.  & 4.2  & 1060 & 1.9  & 7.4\\
17+6  &  4 & 128 & 36 & 0.4 & 11.7 &  330 & 1.8  & 7.2\\

\hline
\end{tabular}
\end{table}

\end{document}